\documentclass[12pt]{article}

\usepackage{amssymb}
\usepackage{amsmath}
\usepackage{amsbsy}
\usepackage{amscd}
\usepackage{amsfonts}
\usepackage{amsthm}
\usepackage{mathrsfs}
\usepackage{verbatim}
\usepackage{hyperref}
\usepackage{fullpage}
\usepackage{mathdots}
\usepackage{graphicx,subfigure}
\usepackage[english]{babel}
\usepackage[utf8]{inputenc}
\usepackage{tikz}
\usepackage{adjustbox}
\usepackage{authblk}
\usepackage{caption}
\usepackage{multicol}

\usepackage{mathtext}
\usepackage{stackengine,scalerel}

\usepackage{tikz-cd}
\usetikzlibrary{backgrounds,fit, matrix}
\usetikzlibrary{positioning}
\usetikzlibrary{calc,through,chains}
\usetikzlibrary{arrows,shapes,snakes,automata, petri}
\usepackage[boxsize=1.25em, centerboxes]{ytableau}

\usetikzlibrary{trees}

\newtheorem{Theorem}[equation]{Theorem}
\newtheorem{Corollary}[equation]{Corollary}
\newtheorem{Lemma}[equation]{Lemma}
\newtheorem{Proposition}[equation]{Proposition}

\theoremstyle{definition}

\newtheorem{Example}[equation]{Example}

\numberwithin{equation}{section}
\numberwithin{figure}{section}

\newcommand{\F}{{\mathbb F}}

\newcommand{\Z}{{\mathbb Z}}

\newcommand{\N}{{\mathbb N}}

\newcommand{\mc}[1]{\mathcal{#1}}

\usepackage[all,2cell]{xy}
	\usepackage{amssymb, hyperref}
	\usepackage[all,2cell]{xy}
	\usepackage{hyperref}
	\usepackage{array}
	\usepackage[capitalise]{cleveref}

\DeclareMathOperator{\Mat}{\mathbf{Mat}}

\begin{document}
	
\title{Evaluation Codes in Bottleneck Metrics}

\author[1]{Mahir Bilen Can}
\author[2]{Dillon Montero}
\author[3]{Ferruh \"Ozbudak}

\affil[1]{\small{Department of Mathematics, Tulane University, USA\\mahirbilencan@gmail.com}}
\affil[2]{\small{Department of Mathematics, Tulane University, USA\\dmontero@tulane.edu}}
\affil[3]{\small{Faculty of Engineering and Natural Sciences, Sabanc{\i} University, Turkey\\ ferruh.ozbudak@sabanciuniv.edu}}
\maketitle

\maketitle

\begin{abstract}
\medskip

Analogs of Reed-Solomon codes are introduced within the framework of bottleneck poset metrics. 
These codes are proven to be maximum distance separable. 
Furthermore, the results are extended to the setting of Algebraic Geometry codes.

\medskip

\medskip
		

\end{abstract}

\section*{Introduction}

Although they are old by modern standards, Reed-Solomon (RS) codes and their variants are still actively used in many new technologies including Cloud Storage systems in junction with other error-correcting codes. 
For example, Amazon S3 Glacier uses RS-like erasure coding to provide durability and reduce the need for replication across multiple servers.
Google Cloud uses similar erasure coding schemes to ensure that data remains recoverable even during server failures.
At the same time, in certain communication systems, data is often structured hierarchically, with varying levels of importance assigned to different parts of a codeword. 
For instance, in multimedia transmissions such as videos or images, certain segments of data, such as headers or control information, are more critical than others, like payload data. 
Errors in these critical components require stricter penalties and more focused correction. 
A poset metric models this hierarchy by defining a partial order, where errors in more significant components, positioned higher in the poset, contribute more heavily to the distance metric. 
Poset-based coding thus enables fine-tuned error correction mechanisms, prioritizing errors in the most important parts of the data.
In this paper, we introduce our variants of the RS codes originally considered in the setting of NRT-metrics,
which have attracted significant attention for their natural structure and applications, particularly in numerical integration techniques, such as discrete Monte Carlo methods. 
Our main contributions are 1) the introduction of a family of RS-like codes in certain subspaces of $s\times r$ matrices with coefficients from the Galois field $\F_q$ with $q$ elements, and 2) extending our newly defined codes to the setting of Algebraic Geometry codes by using function fields. 

\medskip

We now sketch the construction of our codes with full details deferred to a later section.
First, we introduce our metrics.

\begin{figure}[htp]
\begin{center}
\begin{tikzpicture}[scale=.75, every node/.style={circle, draw, fill=white, inner sep=2pt}]    
    \begin{scope}[xshift=-5cm]
    \node (1) at (-2,0) {1};
    \node (2) at (0,0) {2};
    \node (3) at (2,0) {3};
    \node (4) at (-2,2) {4};
    \node (5) at (0,2) {5};
    \node (6) at (2,2) {6};
    \node (7) at (-2,4) {7};
    \node (8) at (0,4) {8};
    \node (9) at (2,4) {9};
    \draw (1) -- (4);
    \draw (2) -- (5);
    \draw (3) -- (6);
    \draw (4) -- (7);
    \draw (5) -- (8);
    \draw (6) -- (9);
    \end{scope}
\begin{scope}[xshift=0]
    \draw [thick, ->, decorate, decoration={zigzag, segment length=4, amplitude=.9}] (-1,2) -- (1,2);
\end{scope}
    \begin{scope}[xshift = 5cm]
    \node (1) at (-2,0) {1};
    \node (2) at (0,0) {2};
    \node (3) at (2,0) {3};
    \node (4) at (0,2) {4};
    \node (5) at (-2,4) {7};
    \node (6) at (0,4) {8};
    \node (7) at (2,4) {9};
    \draw (1) -- (4);
    \draw (2) -- (4);
    \draw (3) -- (4);
    \draw (4) -- (5);
    \draw (4) -- (6);
    \draw (4) -- (7);
    \end{scope}
\end{tikzpicture}
\end{center}
\caption{Constructing the bottleneck poset $U(3,3,2)$ by using $C(3,3)$.}
\label{F:I}
\end{figure}
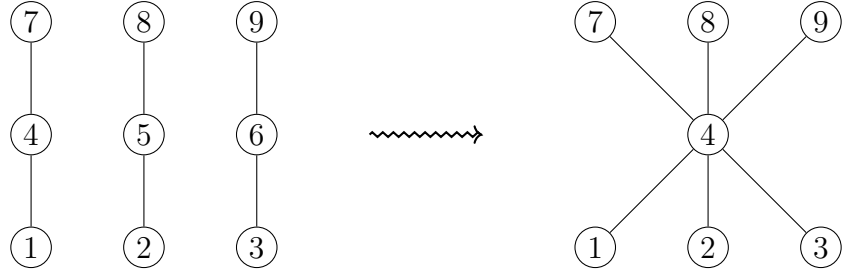

Let $ C(s, r) $ denote the union of $r$ disjoint chains, each of length $ s $. 
We view $ C(s, r) $ as a ranked poset with $r$ connected subposets. 
The main construction of this paper builds on the poset $ (U(s, r, b), \leq) $, which is derived from $ C(s, r) $ by collapsing all vertices of rank $s-b+1$ into a single vertex.
This process is illustrated in Figure~\ref{F:I}.
We will refer to $ U(s, r, b) $ as a \emph{bottleneck poset} since its Hasse diagram resembles the shape of a bottleneck.

Let ${\Mat}_{s\times r}(\F_q)$ denote the vector space of $s\times r$ matrices with entries from $\F_q$. 
For $b\in \{1,\dots, s\}$, we define  
\begin{align}\label{A:bMat}
{\Mat}_{s\times r}^{(b)}(\F_q) := \{ A \in {\Mat}_{s\times r}(\F_q) \mid \text{the $b$-th row of $A$ is constant}\}.
\end{align}
Then, for every element $A\in {\Mat}_{s\times r}^{(b)}(\F_q)$, by identifying the whole $b$-th row of $A$ with an entry in that row, we obtain a vector $A'$ of length $sr-r+1$ whose entries are organized in the shape of the vertices of the Hasse diagram of $U(s,r,b)$ where all edges are removed.
In particular, this process gives a bijection between the set of vertices of $U(s,r,b)$ and the set of distinct coordinates of ${\Mat}_{s\times r}^{(b)}(\F_q)$.
For $x\in U(s,r,b)$, let $A'(x)$ denote the corresponding entry of $A'$.
In this notation, we define the weight of $A$ as follows: 
\begin{align*}
\omega_{U(s,r,b)} (A) = |\{ y\in U(s,r,b)\mid \text{there exists $x\in U(s,r,b)$ is s.t. $y\leq x$ and $A'(x)\neq 0$}\} |.
\end{align*}
As we will show later $\omega_{U(s,r,b)}$ defines a metric on the vector space ${\Mat}_{s\times r}^{(b)}(\F_q)$,
$$
d_{U(s,r,b)} : {\Mat}_{s\times r}^{(b)}(\F_q)\times {\Mat}_{s\times r}^{(b)}(\F_q) \longrightarrow \N.
$$
\medskip

Let $\Z_+$ denote the set of positive integers. 
For $t\in \Z_+$, let $\mathcal{P}(t-1)$ denote the $t$-dimensional $\F_q$-vector space of polynomials of degree at most $t-1$ with coefficients from $\F_q$. 
Let $\alpha_1,\dots, \alpha_r$ be $r$ distinct points from $\F_q$. 
Using these points as evaluation points, to each polynomial $f\in \mathcal{P}(t-1)$ we associate an $s\times r$ matrix, 
$$
H(f;\alpha_1,\dots, \alpha_r) := (\partial^{i-1} f(\alpha_j))_{\substack{i=1,\dots, s \\ j =1,\dots, r}}\ \ \in\ \ {\Mat}_{s\times r}(\F_q), 
$$
where $\partial^{i-1} f(\alpha_j)$ is the value of the $(i-1)$-th hyperderivative of $f$ evaluated at $\alpha_j$.
Notice that if $s=1$, then $H(f;\alpha_1,\dots, \alpha_r)$ is just the row vector $(f(\alpha_1),\dots, f(\alpha_r))$.
In that case, the set 
\begin{align}\label{A:generalizedRS}
RS_{\alpha_1,\dots, \alpha_r}(t-1):=\{ H(f;\alpha_1,\dots, \alpha_r)\mid f\in \mathcal{P}(t-1)\}
\end{align}
is the classical RS code.
For $s>1$, these vector spaces were originally considered by Niederreiter and Pirsic~\cite{NiederreiterPirsic} in the context of digital nets, 
and by Skriganov~\cite{Skriganov2001} in the context of codes.
The objective of this article is to introduce related families of codes that exhibit comparable or enhanced properties depending on their intended applications, for each $b\in \{1,\dots, s\}$.
The first main result of our paper is the following theorem.

\begin{Theorem}\label{T:intro1}
Let $\{\alpha_1,\dots, \alpha_r\}$ be a set of $r\geq 2$ distinct points from $\F_q$.
Let $b\in \N$, $s,t\in \Z_+$. 
Assume that $b\in \{0,\dots, s-1\}$ and $rb+1\leq t\leq rs$.
We set 
$$
\mathcal{P}(t-1)^{(b)}_{\alpha_1,\dots, \alpha_r}:= \{ f\in \mathcal{P}(t-1) \mid \partial^{b} f(\alpha_1) = \cdots = \partial^{b} f(\alpha_r)\}.
$$
If $RS^{s,(b)}_{\alpha_1,\dots, \alpha_r}(t-1)$ denotes the image of the hyperderivative-evaluation map, 
\begin{align*}
{\rm ev}^{(b)}_{\alpha_1,\dots, \alpha_r} : \mathcal{P}(t-1)^{(b)}_{\alpha_1,\dots, \alpha_r}  &\longrightarrow \Mat_{s\times r}^{(b)}(\F_q) \\
f &\longmapsto  
H(f;\alpha_1,\dots, \alpha_r),
\end{align*}
then $RS^{s,(b)}_{\alpha_1,\dots, \alpha_r}(t-1)$ is an $(t-r+1)$-dimensional code.
Furthermore, with respect to bottleneck metric, the minimum distance of $RS^{(b)}_{\alpha_1,\dots, \alpha_r}(t-1)$ is given by 
$$
d_{U(s,r,b)}(RS^{s,(b)}_{\alpha_1,\dots, \alpha_r}(t-1))= rs- t +1.
$$
In particular, $RS^{s,(b)}_{\alpha_1,\dots, \alpha_r}(t-1)$ is an MDS $U(s,r,b)$-code in $\Mat_{s\times r}^{(b)}(\F_q)$.
\end{Theorem}

The second main result of our article is a generalization of our previous theorem to AG codes (to the function field codes to be more precise). 
To explain, let us consider a global function field $F/\F_q$ with full constant field $\F_q$. 
In this article only rational places of $F$, although our theorems can be extended to codes constructed with higher degree places as in~\cite{NiOz2002}. 
Additionally, to simplify the notation, we will consider the bottleneck metric codes with respect to the poset $U(s,r,0)$ only.

For a rational place $P$ of $F$, let $\nu_P$ denote the corresponding normalized discrete valuation of $F$. 
We fix $r$ mutually distinct places $P_1,\dots, P_r$ in $F$, and let $G$ be a divisor of $F/\F_q$. 
For each $i\in \{1,\dots, r\}$, let $n_i$ denote the coefficient of $P_i$ in $G$. 
For a rational function $f$ from the Riemann-Roch space $\mc{L}(G)$, let $-a_i$ denote the value $\nu_{P_i}(f)$. 
If we consider the unique local expression $f= \sum_{j=-\infty}^\infty c_{i,j}t_i^j$, where $t_i$ is the local parameter of $P_i$ in $F$, then $-a_i$ is the coefficient of the smallest nonzero term in the expression. 
We associate to $f$ the following $s\times r$ matrix:
\begin{align*}
f \longmapsto
\mathbf{c}(f) := 
\begin{bmatrix}
c_{1,-n_1} & c_{2,-n_2} & \cdots & c_{r,-n_r} \\ 
c_{1,-n_1+1} & c_{2,-n_2+1} & \cdots & c_{r,-n_r+1} \\ 
\vdots & \vdots & \ddots & \vdots \\
c_{1,-n_1+s-1} & c_{2,-n_2+s-1} & \cdots & c_{r,-n_r+s-1}
\end{bmatrix}.
\end{align*}
In this notation, the following subspace of the Riemann-Roch space associated with $G$ plays a crucial role in the second main theorem of our paper:  
\[
\mathcal{L}^{(0)}_{P_1, \dots, P_r}(G) := \{ f \in \mathcal{L}(G) \mid c_{1,-n_1}(f) = \cdots = c_{r,-n_r}(f) \},  
\]  
where $c_{i,-n_i}(f)$ denotes the coefficient of the lowest-degree term in the local expansion  
\[
\sum_{j=-n_i}^\infty c_{i,j} t_i^j  
\]  
of $f$ at $P_i$, for $i \in \{1, \dots, r\}$.

\begin{Theorem}\label{T:MDSURT}
Let $F/\F_q$ be a global function field of genus $g\geq 1$.
Let $r$, $k$ and $s$ be positive integers satisfying the inequalities 
\begin{itemize}
\item $2r+s-g\leq rs$,
\item $0\leq k-1\leq g\leq r$, and 
\item $g-1\leq s$.
\end{itemize}
We assume that $F/\F_q$ has at least $r$ rational places.
Let $h$ denote the divisor class number of $F$. 
Finally, let $A_k$ denote the number of positive divisors of $F$ degree $k$. 
If the inequality 
\begin{align}\label{A:keyinequality}
{ r+s + k - g \choose r-1 } A_k < h
\end{align}
holds, then the parameters of the AG code defined by $\mathcal{U} := \{ \mathbf{c}(f) \mid f\in \mathcal{L}^{(0)}_{P_1,\dots, P_r}(G) \}$
satisfy the following inequalities 
\begin{enumerate}
\item[(1)] $d_{U(s,r,0)}(\mathcal{U}) \geq s-g+2+k$,  
\item[(2)] $\dim \mathcal{U} \geq rs - r +1 -s$.
\end{enumerate}
\end{Theorem}

As a direct consequence of our previous theorem, we obtained the following result.

\begin{Corollary}\label{C:MDSURT}
We maintain the notation from the previous theorem.
If the genus of $F$ is $k-1$, then $\mathcal{U}$ is an MDS bottleneck metric code. 
In particular, if $F$ is the function field of an elliptic curve, then $\mathcal{U}$ is either an MDS bottleneck-metric code or a near MDS bottleneck metric code.
\end{Corollary}

The structure of our paper is as follows. 
In the next section, we introduce our basic objects, including poset metrics and discuss NRT codes.
Additionally, we introduce the Reed-Solomon codes in this context. 
In Section~\ref{S:RSinBottleneck} we prove our first main result, Theorem~\ref{T:intro1}. 
In Section~\ref{S:AGinBottleneck} we prove our second main result and its corollary.

\section{Preliminaries and Notation}\label{S:Preliminaries} 

We begin with reviewing the theory of general poset metrics, which were first introduced in~\cite{BGL}. 

\subsection*{Poset metrics.}
Let $n\in \Z_+$. 
Let $(P,\leq)$ be a poset whose underlying set is given by $[n]:=\{1,\dots, n\}$. 
We will use $P$ to define a metric on the vector space $\F_q^n$.
First, the {\em support} of a vector $v=(v_1,\dots, v_n) \in \F_q^n$ is defined by  
$$
{\rm supp}(v) := \{ i\in \{1,\dots, n\} \mid v_i \neq 0 \}.
$$ 
Next, we consider the {\em lower order ideal generated by ${\rm supp}(v)$}:
$$
P_v:= \{ j\in [n] \mid j \leq i \text{ for some $i\in {\rm supp}(v)$}\}.
$$
Then the {\em $P$-weight} of $v$ is defined by 
$$
\omega_P(v) := |P_v |.
$$
The {\rm $P$-metric on $\F_q^n$} is the metric defined by 
\begin{align*}
d_P : \F_q^n \times \F_q^n &\longrightarrow \N \\
(v,w) &\longmapsto \omega_P(v-w).
\end{align*}
Let $C$ be a subset of $\F_q^n$. 
We denote the minimum nonzero weight $\omega_P(v)$, where $v\in C$, by $d_P(C)$. 
In this notation, the analog of the Singleton's bound for poset metric $d_P$ is as follows: 
\begin{align}\label{A:Singletonforposets}
|C| \leq q^{n - d_P(C) +1}.
\end{align}
In particular, if $C$ is a linear $[n,k]_q$ $P$-code, then we have 
\begin{align*}
d_P(C) \leq n-k+1.
\end{align*}
A proof of this result can be found in~\cite[Proposition 2.1]{HyunKim}.

\subsection*{A Special case: the NRT-metric.}

Let $\mathbf{a}:=(a_1,\dots, a_s)\in \F_q^s$.
We will view $\mathbf{a}$ as one of $r$ equal-length segments of a codeword.
The weight of each segment is defined as follows: 
$$
\omega (\mathbf{a}) = 
\begin{cases}
0 & \text{ if $\mathbf{a} = (0,\dots, 0)$, } \\
\max\{j \mid a_j \neq 0\} &  \text{ if $\mathbf{a} \neq (0,\dots, 0)$. }
\end{cases}
$$
To make this weight easier to remember, we convert $\mathbf{a}$ into a column matrix by taking its reverse-transpose: 
$$
A:=
\begin{bmatrix}
a_s \\
a_{s-1}\\
\vdots \\
a_1
\end{bmatrix}.
$$
In this notation, the weight of a nonzero column vector $A$ is given by $s- i+1$, where $i$ is index of the first nonzero row of $A$. 
For example, if $\mathbf{a}$ is the vector $\mathbf{a}:=(0,1,1,0,1,0,0,0)$, then the weight of $A$ is 5 since the last nonzero entry of $\mathbf{e}$ is at the 5th position. Equivalently, we have 
$$
\omega \left(
\begin{bmatrix}
0 \\ 0 \\ 0 \\ 1 \\ 0 \\ 1\\ 1\\ 0
\end{bmatrix}
\right) = 8 - 4 +1 = 5.
$$

The poset metric interpretation of this weight is obtained as follows.
Let $r,s\in \Z_+$ be such that $n = rs$.
Recall that $C(s,r)$ denotes the union of $r$ disjoint chains, each containing $s$ vertices. 
We label the vertices of $C(s,r)$ from 1 to $n$ as follows:
\begin{enumerate}
\item Label the smallest elements of each chain by using the numbers $1,\dots, r$ starting from the left most chain towards right. 
\item Move to the next level and repeat.
\end{enumerate}
For example, in Figure~\ref{F:C(2,3)}, we depict $C(2,3)$ with vertices labeled just as defined. 
\begin{figure}[htp]
\begin{center}
\begin{tikzpicture}[scale=.8, every node/.style={circle, draw, fill=white, inner sep=2pt}]
    \node (1) at (0,0) {1};
    \node (2) at (0,1.5) {4};
    \node (3) at (1,0) {2};
    \node (4) at (1,1.5) {5};
    \node (5) at (2,0) {3};
    \node (6) at (2,1.5) {6};
    \draw (1) -- (2);
    \draw (3) -- (4);
    \draw (5) -- (6);
\end{tikzpicture}
\end{center}
\caption{The Hasse diagram of $C(2,3)$.}
\label{F:C(2,3)}
\end{figure}

By using our previous discussion, we now define the NRT-metric on $\Mat_{s\times r}(\F_q)$. 
Initially, let $A$ be a column vector with $s$ entries. 
As before, let $i$ denote the index of the first nonzero entry of $A$. 
Then we set, 
\begin{align*}
\omega_{C(s,1)}(A) :=  s-i+1,
\end{align*}
Next, let us assume that $A$ is an $s\times r$ matrix, $A:=(a_{ij})_{i=1,\dots,s \atop j=1,\dots, r}\in \Mat_{s\times r}(\F_q)$.
Let $A_1,\dots, A_r$ denote the columns of $A$. 
Then the {\em NRT-weight} of the matrix $A$, denoted $\omega_{C(s,r)}(A)$, is defined as 
$$
\omega_{C(s,r)}(A):= \omega_{C(s,1)}(A_1)+\cdots + \omega_{C(s,1)}(A_r). 
$$
It is easy to check that this sum is the value of the $P$-weight of $A$ where $P$ is the poset $C(s,r)$. 
Hereafter, we denote the corresponding NRT-metric by $d_{C(s,r)}$.

\subsection*{Bottleneck metrics.}

At the beginning of our article, we introduced our bottleneck metrics.
In this section we provide more details on these metrics. 
\medskip

First, let us introduce a notion of a rank function. 
Let $P$ be a finite poset. 
Let $x$ and $y$ be two distinct elements from $P$. 
The element $x$ is said to cover $y$ if $y\leq x$ and there is no other element $z\in P\setminus \{x,y\}$ such that $y\leq z \leq x$. 
In this case, we call the pair $(y,x)$ a covering relation. 
The integer valued function $\ell : P\to \N$ such that 
\begin{itemize}
\item $\ell(x) = \ell(y) +1$ for every covering relation $(y,x)$, 
\item $\ell(z) = 0$ for every minimal element $z\in P$,
\end{itemize}
is called the rank function of $P$. 
Then the rank of an element $x\in P$ is defined to be $\ell(z)$.

It is easy to see that not every poset possesses a rank function. 
If there exists one, then call $P$ a ranked poset. 
Evidently the disjoint union of chains $C(s,r)$ is a ranked poset. 
Our bottleneck posets $U(s,r,b)$ are also ranked posets. 
Furthermore, in both $C(s,r)$ and $U(s,r,b)$ every maximal element has the same rank, $s-1$.
It is also evident that $U(s,r,b)$ has exactly $sr- r +1$ vertices. 

We will maintain the labeling of the vertices of $U(s,r,b)$ induced from the labeling of the vertices $C(s,r)$ which we described in the previous subsection. 
Since the elements of rank $s-b+1$ in $C(s,r)$ are identified with each other, we will use the following numbers: 
$$
1,\dots, (s-b)r,\ (s-b)r+1,\ \underbrace{ (s-b)r+2,\dots , (s-b)r+r}_{\text{omitted}}, \ (s-b)r+r+1, \dots, sr. 
$$
In Figure~\ref{F:U(3,3,2)}, we depict $U(3,3,2)$ with vertices labeled just as defined. 
\begin{figure}[htp]
\begin{center}
\begin{tikzpicture}[scale=.75, every node/.style={circle, draw, fill=white, inner sep=2pt}]    
    \begin{scope}[xshift = 5cm]
    \node (1) at (-2,0) {1};
    \node (2) at (0,0) {2};
    \node (3) at (2,0) {3};
    \node (4) at (0,2) {4};
    \node (5) at (-2,4) {7};
    \node (6) at (0,4) {8};
    \node (7) at (2,4) {9};
    \draw (1) -- (4);
    \draw (2) -- (4);
    \draw (3) -- (4);
    \draw (4) -- (5);
    \draw (4) -- (6);
    \draw (4) -- (7);
    \end{scope}
\end{tikzpicture}
\end{center}
\caption{The Hasse diagram of the bottleneck poset $U(3,3,2)$.}
\label{F:U(3,3,2)}
\end{figure}
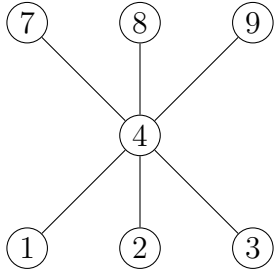 

We will now discuss the relationship between the metrics $d_{U(s,r,b)}$ and $d_{C(s,r)}\big\vert_{\small{\Mat_{s\times r}^{(b)}(\F_q)\times \Mat_{s\times r}^{(b)}(\F_q)}}$.
It is notationally more convenient to work with the corresponding weights $\omega_{U(s,r,b)}$ and $\omega_{C(s,r)}$, respectively. 
First, we show by an example that these weight functions take different values. 
\begin{Example} 
Let $A \in \Mat_{3\times 3}(\F_q)$ be the matrix defined by 
$$
A = 
\begin{bmatrix}
1 & 0 & 1 \\
1 & 1 & 1 \\
0 & 0 & 1 
\end{bmatrix}.
$$
Then we have the weights 
$$
\omega_{C(3,3)}(A) = 8\quad \text{ and } \quad \omega_{U(3,3,2)}(A) = 6.
$$
\end{Example}

Next, we will analyze the matrices on which our weight functions agree. 

\begin{Lemma}\label{L:theyagree}
Let $A$ be an element of $\Mat_{s\times r}^{(b)}(\F_q)$.
Let $A_0$, $A_1$, and $A_2$ be the matrices obtained from $A$ by removing its last $s-b+1$ rows, last $s-b$ rows, and the first $b$ rows, respectively. 
Then the following assertions hold: 
\begin{enumerate}
\item[(1)] if $A_1= 0$, then $\omega_{U(s,r,b)}(A) = \omega_{C(s,r)}(A)$;
\item[(2)] if $A_0=0$ but $A_1\neq 0$, then $\omega_{U(s,r,b)}(A) = (s-b)r+1$;
\item[(3)] if $A_0\neq 0$, then $\omega_{U(s,r,b)}(A) = \omega_{C(b-1,r)}(A_0) + (s-b)r+1$.
\end{enumerate}
\end{Lemma}

\begin{proof}
(1) If $A_1$ is the zero matrix in $\Mat_{b\times r}(\F_q)$, then the $\omega_{U(s,r,b)}$-weight of $A$ is solely determined by the entries below the $b$-th row. 
Since the relations of the posets $U(s,r,b)$ and $C(s,r)$ are identical in their portions below rank $b$, the weights $\omega_{U(s,r,b)}$ and $\omega_{C(s,r)}$ 
are the same in that range. 
This proves (1). 

(2) If $A_0=0$ but $A_1\neq 0$, then the $b$-th row of $A$ is nonzero but all earlier rows are zero. 
In this case, by the definition of the $U(s,r,b)$-metric, the weight of $A$ is given by the number of vertices of $U(s,r,b)$ of rank at most $s-b+1$. 
Since there are exactly $(s-b)r+1$ such vertices, we have $\omega_{U(s,r,b)}(A) = (s-b)r+1$. 

(3) If $A_0\neq 0$, then there a nonzero entry of $A$ corresponding entry in $U(s,r,b)$ has rank greater than $s-b+1$. 
It follows that all vertices of $U(s,r,b)$ with rank at most $s-b+1$ contribute to the $\omega_{U(s,r,b)}$-weight of $A$. 
At the same time, the portion of the poset $U(s,r,b)$ above rank $s-b+1$ is identical with the poset $C(b-1,r)$.
It follows that $\omega_{U(s,r,b)}(A) = \omega_{C(b-1,r)}(A_0) + (s-b)r+1$.
This finishes the proof of our lemma. 
\end{proof}

\subsection*{Reed-Solomon Codes in the Hamming metric.}

Classical Reed-Solomon code of length $n$ dimension $k$ is an evaluation code, defined as follows.
Let $\mathcal{P}(k-1)$ denote the $\F_q$-vector space consisting of polynomials $f(z)\in \F_q [ z]$ such that $\deg (f) \leq k-1$. 
Hence, we have 
$$
\dim \mathcal{P}(k-1) = k.
$$
Let $\alpha_1,\dots, \alpha_n$ be $n$ distinct points from the affine line $\mathbb{A}^1(\F_q)$. 
The associated Reed-Solomon code, denoted by $RS_{\alpha_1,\dots, \alpha_r}(t-1)$, is given by the image of the evaluation map 
\begin{align*}
{\rm ev}_{\alpha_1,\dots, \alpha_n} : \mathcal{P}(k-1) &\longrightarrow \F_q^n \\
f(z) &\longmapsto (f(\alpha_1),\dots, f(\alpha_n)).
\end{align*}
It is well-known that the Reed-Solomon code meets the upper bound in the Singleton's theorem:
$$
\dim_{\F_q} RS_{\alpha_1,\dots, \alpha_n}(t-1) = n- d_H(RS_{\alpha_1,\dots, \alpha_n}(t-1)) +1,
$$
where $d_H$ is the ordinary Hamming metric defined on $\F_q^n$.

\subsection*{Reed-Solomon Codes in the NRT-metric.}

There is a natural analog of a Reed-Solomon code in the setting of the NRT-metric. 
It is constructed as follows. 
Let $f(z)\in \mathcal{P}(t-1)$.
Let $f_0,\dots,f_{t-1}$ denote the coefficients of $f(z)$ in the basis $1,z,\dots, z^{t-1}$:
\begin{align}\label{A:standardexpansion}
f(z) = f_0 + f_1 z + \cdots + f_{t-1} z^{t-1}.
\end{align}
The {\em $j$-th hyperderivative} of $f(z)$ is the polynomial defined by
\begin{align}\label{A:hyperderivative}
\partial^j f(z) := {0\choose j} f_0 z^{-j} + {1\choose j} f_1 z^{1-j} + \cdots + {t-1\choose j} f_{t-1} z^{t-1-j}, 
\end{align}
where we use the convention that 
$$
{i\choose j} := 
\begin{cases}
\frac{ i ! }{ (i-j)! j! } & \text{ if\ $0\leq j \leq i$,} \\
0 & \text{ otherwise.}
\end{cases}
$$
Notice that 
$$
\partial^j f(z)=0\quad \text{ if \quad $j > \deg(f)$}.
$$

\bigskip

Let $\alpha \in \mathbb{A}^1(\F_q)$. 
Then the {\em Taylor expansion of $f(z)$ at $\alpha$} is given by 
\begin{align}\label{A:Taylor}
f(z) = \sum_{j=0}^{t-1} \partial^j f(\alpha) (z-\alpha)^j.
\end{align}
In particular, $\alpha$ is a root of $f(z)$ with multiplicity $m$ if and only if 
$\partial^j f(\alpha) = 0$ for $j\in \{0,1,\dots, m-1\}$ and $\partial^m f(\alpha) \neq 0$. 
\bigskip

There is a useful product rule for the hyperderivatives. 
\begin{Lemma}(\cite[Lemma 6.47]{NiederreiterLidl})\label{L:productrule}
For polynomials $f_1,\dots, f_t \in K[z]$, we have
$$
\partial^n (f_1\cdots f_t) = \sum_{n_1,\dots, n_t \geq 0\atop n_1+\cdots +n_t = n} (\partial^{n_1}f_1) \cdots (\partial^{n_t} f_t).
$$
\end{Lemma}

An immediate consequence of Lemma~\ref{L:productrule} is that if $f(z)$ is given by $(z-c)^t$ for some $c\in K$ and $t\in \Z$, 
then we have 
\begin{align}\label{A:productrule}
\partial^n f = {t\choose n} (z-c)^{t-n}.
\end{align}
In particular, for $n=t$, we have $\partial^t f(z) = 1$, while for $n>t$ we have $\partial^t f(z)$  is 0.
\bigskip

We now introduce the NRT-metric analogs of the Reed-Solomon codes, which are constructed by Skriganov in~\cite{Skriganov2001}.

\begin{Theorem}\label{T:NRTRS}
Let $r\in \{1,\dots, q\}$. 
Let $\alpha_1,\dots, \alpha_r$ denote $r$ distinct points from $\mathbb{A}^1(\F_q)$.
Let $s$ and $t$ be two positive integers such that $t\leq rs$.
Let $RS_{\alpha_1,\dots, \alpha_r}^{s}(t-1)$ denote the code defined as the image of the following evaluation map: 
\begin{align*}
{\rm ev}_{\alpha_1,\dots, \alpha_r} : \mathcal{P}(t-1) &\longrightarrow \Mat_{s\times r}(\F_q) \\
f(z) &\longmapsto 
\begin{bmatrix}
f(\alpha_1) & f(\alpha_2) & \cdots & f(\alpha_r) \\
\partial^{ 1 } f(\alpha_1) & \partial^{ 1 } f(\alpha_2) & \cdots & \partial^{ 1 } f(\alpha_r) \\
\vdots & \vdots & \ddots & \vdots \\
\partial^{ s-2 } f(\alpha_1) & \partial^{ s-2 } f(\alpha_2) & \cdots & \partial^{ s-2 } f(\alpha_r) \\
\partial^{ s-1 } f(\alpha_1) & \partial^{ s-1 } f(\alpha_2) & \cdots & \partial^{ s-1 } f(\alpha_r) \\
\end{bmatrix}.
\end{align*}
Then $RS_{\alpha_1,\dots, \alpha_r}^{s}(t-1)$ is an $[n,t]_q$ MDS NRT-metric code, where $n=rs$. 
\end{Theorem} 
A proof of the previous theorem can be found in~\cite[Theorem 4.4]{FirerAlvesPinheiroPanek}.
Caution: the authors in~\cite{FirerAlvesPinheiroPanek} work with the transpose of the image of ${\rm ev}_{\rm NRT}$ that we introduced here.
\medskip

\subsubsection*{Orders of vanishing.}

We conclude our preparatory section by a brief discussion of the relationship between the NRT weights and the orders of vanishing of the polynomials.
\medskip

Let $f\in \mathcal{P}(t)$. 
The {\em order of vanishing} of $f$ at a point $\alpha$ from $\F_q$ will be denoted by $\nu_f(\alpha)$.
In particular, we have 
$$
\nu_f(\alpha)=0 \implies f(\alpha)\neq 0.
$$
In relation with this observation, we state a lemma whose proof is simple, hence, omitted. 
\begin{Lemma}
Let $A$ denote the matrix ${\rm ev}_{\alpha_1,\dots, \alpha_r}(f)$.
If $A_j$ denotes the $j$-th column of $A$, then the NRT-weight of $A_j$ is given by 
$$
\omega_{C(s,1)}(A_j) = s-\nu_f(\alpha_j).
$$
Therefore, the NRT-weight of the matrix $A$ is given by 
\begin{align}\label{A:onMatsr}
\omega_{C(s,r)}(A) = \sum_{j=1}^r (s-\nu_f(\alpha_j)) = rs  - \sum_{j=1}^r \nu_f(\alpha_j).
\end{align}
\end{Lemma}

\section{Reed-Solomon Codes in the Bottleneck Metrics}\label{S:RSinBottleneck}

In the preparatory section, we reviewed Reed-Solomon codes in the Hamming and NRT metrics.
Here, we initiate the construction of our analogs of Reed-Solomon codes, starting with the bottleneck metric on matrices.
We begin with presenting a key calculation for our first main theorem.

Let $\alpha_1,\dots, \alpha_r$ be $r$ distinct points from $\mathbb{A}^1(\F_q)$. 
Let $s\in \N$.
Recall that $\mc{P}(t)^{(b)}_{\alpha_1,\dots, \alpha_r}$ denotes the polynomials of degree at most $t$ and whose $s$-th hyperderivatives 
evaluate to the same constant,
$$
\mathcal{P}(t)^{(s)}_{\alpha_1,\dots, \alpha_r}:= \{ f(z) \in \mathcal{P}(t) \mid 
\partial^{s} f(\alpha_1) = \partial^{s} f(\alpha_2) = \cdots = \partial^{s} f(\alpha_r) \}.
$$

\begin{Lemma}\label{L:useful2}
Let $\alpha_1,\dots, \alpha_r$ be $r$ distinct points from $\mathbb{A}^1(\F_q)$. 
Let $s$ and $t$ be two positive integers.
Then 
$\mathcal{P}(t)^{(s)}_{\alpha_1,\dots, \alpha_r}$ is a vector space.
Furthermore, we have 
$$
\dim\mathcal{P}(t)^{(s)}_{\alpha_1,\dots, \alpha_r} =
\begin{cases}
t+1 & \text{ if $s>t$,} \\
t-r+2 & \text{ if $s\leq t$.}
\end{cases} 
$$
\end{Lemma}

\begin{proof}
The fact that $\mathcal{P}(t)^{(s)}_{\alpha_1,\dots, \alpha_r}$ is a vector space is easy to check. 
We proceed to calculate its dimension.

Clearly, if $t< s$, then the $s$-th hyperderivative of any element of $\mathcal{P}(t)$ is identically 0.
Therefore, in this case, we have $\mathcal{P}(t)^{(s)}_{\alpha_1,\dots, \alpha_r}=\mathcal{P}(t)$,
implying that $\dim\mathcal{P}(t)^{(s)}_{\alpha_1,\dots, \alpha_r}= \dim \mathcal{P}(t) = t+1$. 
We proceed with the assumption that $s \leq t$.

Let $f(z)\in \mathcal{P}(t)$ be given by $f(z):= x_0 + x_1 z+\cdots + x_tz^t$ for some $x_0,\dots, x_t$ from $\F_q$. 
We impose the conditions 
\begin{align}\label{A:imposedforsomec}
\partial^{s} f(\alpha_1) = \partial^{s} f(\alpha_2) = \cdots = \partial^{s} f(\alpha_r)=c,
\end{align}
where $c\in \F_q$.
These equations can be written as a system of equations in the variables $x_0,x_1,\dots, x_t$:
\begin{align*}
 {0\choose s} x_0 \alpha_1^{-s} + {1\choose s} x_1 \alpha_1^{1-s} + \cdots + {t\choose s} x_{t} \alpha_1^{t-s} &= c \\
 {0\choose s} x_0 \alpha_2^{-s} + {1\choose s} x_1 \alpha_2^{1-s} + \cdots + {t\choose s} x_{t} \alpha_2^{t-s} &= c \\
 &\vdots \\
 {0\choose s} x_0 \alpha_r^{-s} + {1\choose s} x_1 \alpha_r^{1-s} + \cdots + {t\choose s} x_{t} \alpha_r^{t-s} &= c 
\end{align*}
Then we write these equations in a matrix form $A X = C$, where $X$ and $C$ 
are the column vectors $X= (x_0,\dots, x_t)^\top$, $C=(c,\dots, c)^\top$,
and $A$ is the matrix given by 
$$
A:=
\begin{pmatrix}
 {0\choose s}  \alpha_1^{-s} & {1\choose s} \alpha_1^{1-s} & \cdots & {t\choose s}  \alpha_1^{t-s}  \\ 
  {0\choose s}  \alpha_2^{-s} & {1\choose s}  \alpha_2^{1-s} & \cdots & {t\choose s}  \alpha_2^{t-s}  \\ 
  \vdots & \vdots & \ddots & \vdots \\
   {0\choose s} \alpha_r^{-m} & {1\choose s}  \alpha_r^{1-s} & \cdots & {t\choose s}  \alpha_r^{t-s} 
\end{pmatrix} = 
\begin{pmatrix}
 0 & \cdots & 0 & {s\choose s}  \alpha_1^{0} & {s+1\choose s} \alpha_1^{1} & \cdots & {t\choose s}  \alpha_1^{t-s}  \\ 
 0 & \cdots & 0 &  {s\choose s}  \alpha_2^{0} & {s+1\choose s}  \alpha_2^{1} & \cdots & {t\choose s}  \alpha_2^{t-s}  \\ 
  \vdots &  & \vdots &  \vdots & \vdots & \ddots & \vdots \\
  0 & \cdots & 0 &  {s\choose s} \alpha_r^{0} & {s+1\choose s}  \alpha_r^{1} & \cdots & {t\choose s}  \alpha_r^{t-s} 
\end{pmatrix}.
$$
Notice that since $\alpha_1,\dots, \alpha_r$ are distinct elements of $\F_q$, and since the nonzero entries of the matrix $A$ can be recognized as a part of the Vandermonde-like matrix, where the binomial coefficients are viewed as nonzero weights, we see that $A$ has linearly independent rows. 
Since the first $s$ columns of $ A $ are zero and the remaining $ r \times (t - s+1) $ submatrix has rank $ r $, $ A $ has the form:
\[
A = \begin{pmatrix} 0_{r \times s} & A_{r \times (t-s+1)} \end{pmatrix},
\]
where $ A_{r \times (m-s)} $ is the nonzero submatrix of $ A $ with full rank $ r $. This means that $ A_{r \times (t-s)} $ is surjective, 
mapping an $(t - s)$-dimensional space to an $ r $-dimensional space.
Thus, the equation $ A X = C $ becomes:
\[
\begin{pmatrix} 0_{r \times s} & A_{r \times (t-s+1)} \end{pmatrix} \begin{pmatrix} X_1 \\ X_2 \end{pmatrix} = C,
\]
where $ X_1 $ is the $ s \times 1 $ subvector corresponding to the first $ s $ entries of $ X $, and $ X_2 $ is the $ (t - s+1) \times 1 $ subvector corresponding to the remaining $t - s +1$ entries.
This simplifies to:
\[
A_{r \times (t-s+1)} X_2 = C \qquad \text{(for some $C:=(c,\dots,c)^\top \in \F_q^{t-s+1}$)}.
\]
Since $ A_{r \times (t-s+1)} $ has full rank $ r $, the number of free variables is seen to be
$$
(t-s+1)-r.
$$
Hence, in total, the number of free variables is $s + (t-s+1)-r = t+1-r$.
Since $c$ varies in $\F_q$, the dimension of the vector space of polynomials $f(z)\in \mathcal{P}(t)$ 
satisfying the equations (\ref{A:imposedforsomec}) for some $c\in \F_q$ is given by $t-r+2$. 
This finishes the proof.
\end{proof}

We are now ready to prove our first main result, Theorem~\ref{T:intro1}.
Let us recall its statement for convenience of the reader. 

\medskip
Let $\{\alpha_1,\dots, \alpha_r\}$ be a set of $r\geq 2$ distinct points from $\F_q$.
Let $b\in \N$, $s,t\in \Z_+$. 
Assume that $b\in \{0,\dots, s-1\}$ and $rb+1\leq t\leq rs$.
We set 
$$
\mathcal{P}(t-1)^{(b)}_{\alpha_1,\dots, \alpha_r}:= \{ f\in \mathcal{P}(t-1) \mid \partial^{b} f(\alpha_1) = \cdots = \partial^{b} f(\alpha_r)\}.
$$
If $RS^{s,(b)}_{\alpha_1,\dots, \alpha_r}(t-1)$ denotes the image of the hyperderivative-evaluation map, 
\begin{align*}
{\rm ev}^{(b)}_{\alpha_1,\dots, \alpha_r} : \mathcal{P}(t-1)^{(b)}_{\alpha_1,\dots, \alpha_r}  &\longrightarrow \Mat_{s\times r}^{(b)}(\F_q) \\
f &\longmapsto  
H(f;\alpha_1,\dots, \alpha_r),
\end{align*}
then $RS^{s,(b)}_{\alpha_1,\dots, \alpha_r}(t-1)$ is an $(t-r+1)$-dimensional code.
Furthermore, with respect to bottleneck metric, the minimum distance of $RS^{(b)}_{\alpha_1,\dots, \alpha_r}(t-1)$ is given by 
$$
d_{U(s,r,b)}(RS^{s,(b)}_{\alpha_1,\dots, \alpha_r}(t-1))= rs- t +1.
$$
In particular, $RS^{s,(b)}_{\alpha_1,\dots, \alpha_r}(t-1)$ is an MDS $U(s,r,b)$-code in $\Mat_{s\times r}^{(b)}(\F_q)$.

\begin{proof}[Proof of Theorem~\ref{T:intro1}]

Our evaluation map is injective since it is given by the restriction to the subspace 
$\mathcal{P}^{(b)}_{\alpha_1,\dots, \alpha_r}(t-1)\subset \mathcal{P}(t-1)$ of the injective evaluation map 
${\rm ev}_{\alpha_1,\dots, \alpha_r} : \mathcal{P}(t-1) \to {\Mat}_{s\times r}(\F_q)$ that defines the NRT-Reed-Solomon code. 
Under our assumptions, by Lemma~\ref{L:useful2}, the subspace $\mathcal{P}^{(b)}_{\alpha_1,\dots, \alpha_r}(t-1)$ is $(t-r+1)$-dimensional. 
Hence, $RS^{s,(b)}_{\alpha_1,\dots, \alpha_r}(t-1)$ is a code of dimension $t-r+1$. 
This proves our first assertion. 
We proceed to prove our second assertion.

First, we produce a candidate polynomial $f$ whose evaluation at the points $\alpha_1,\dots, \alpha_r$ gives the right minimum distance. 
Later, we will prove that the $U(s,r,b)$-weight of our candidate is indeed the desired minimum distance. 
We begin with an MDS NRT code of a small size. 

By Theorem~\ref{T:NRTRS}, we know that the evaluation code 
$$
{\rm ev}_{\alpha_1,\dots, \alpha_r}(\mathcal{P}(t-rb-1)) \subset \Mat_{(s-b)\times r}(\F_q)
$$ 
is an MDS NRT-code.
Hence, it is minimum distance is given by 
$$
\text{length} -\text{dimension} + 1 = r(s-b) - (t-rb)+1 = rs - t +1. 
$$
Now let $g(z)\in \mathcal{P}(t-rb-1)$ denote the polynomial such that the NRT-weight of ${\rm ev}_{\alpha_1,\dots, \alpha_r}(g) \in \Mat_{(s-1) \times r}(\F_q)$ is 
given by 
$$
\omega_{C(s-b,r)}({\rm ev}_{\alpha_1,\dots, \alpha_r}(g)) = rs-t+1.
$$
In other words, ${\rm ev}_{\alpha_1,\dots, \alpha_r}(g)$ is a minimum weight codeword for the NRT-metric on $\Mat_{(s-b) \times r}(\F_q)$. 
Thus, by (\ref{A:onMatsr}), we see that 
$$
rs-t+1= r(s-b) - \sum_{j=1}^r \nu_g(\alpha_j).
$$
Hence, we have 
\begin{align}\label{A:forg}
\sum_{j=1}^r \nu_g(\alpha_j) = t-rb-1
\end{align}
Let $f(z)$ be the polynomial defined by $f(z) := g(z) \prod_{m=1}^{r} (z- \alpha_m)^b$, 
where $g(z)$ is as before.
Since $f(z)$ is obtained from $g(z)$ by adding the distinct roots $\alpha_1,\dots, \alpha_r$ each repeated $b$ times, we have 
\begin{align}\label{A:fandg}
\nu_f(\alpha_i) = b+ \nu_g(\alpha_i).
\end{align}
Hence, combined with (\ref{A:forg}) we see that 
\begin{align}\label{A:forf}
\sum_{j=1}^r \nu_f(\alpha_j) = t-1
\end{align}
We proceed to calculate the $U(s,r,b)$-weight of $f$. 
To ease our notation, we set 
$$
A:= {\rm ev}_{\alpha_1,\dots, \alpha_r}(f).
$$
Let $A_1$ be the matrix obtained from $A$ by removing its last $s-b$ rows.
Since the multiplicity of each $\alpha_j$ ($j\in \{1,\dots, r\}$) as a root of $f$ is at least $b$, we see that $A_1=0$. 
Then, by Lemma~\ref{L:theyagree} (1), the $U(s,r,b)$-weight of $A$ is the NRT-weight $\omega_{C(s,r)}(A)$. 
By using (\ref{A:onMatsr}) once more together with (\ref{A:forf}), we obtain the $U(s,r,b)$-weight of $A$: 
\begin{align*}
\omega_{U(s,r,b)}(A) &= \omega_{C(s,r)}(A) \\
&= rs - \left(\sum_{j=1}^r \nu_f(\alpha_j)\right) \\
&= rs- (t-1)\\
&= rs-t+1.
\end{align*}
Now, since $rb+1 \leq t$, we know that $\omega_{U(s,r,b)}(A) = rs-t+1\leq rs-rb=r(s-b)$. 
We will now show that $A$ is indeed a minimum weight codeword.  

Let $B$ be another codeword from $RS^{s,(b)}_{\alpha_1,\dots, \alpha_r}(t-1)$ that comes from a polynomial $h\in \mc{P}^{(b)}_{\alpha_1,\dots,\alpha_r}(t-1)$. 
Let $B_1$ denote the matrix obtained from $B$ by removing its last $s-b$ rows.
If $B_1\neq 0$, then by Lemma~\ref{L:theyagree} we know that the $U(s,r,b)$-weight of $B$ is at least $r(s-b)+1$. 
This means that if a codeword $B$ of $RS^{s,(b)}_{\alpha_1,\dots, \alpha_r}(t-1)$ has the smallest nonzero weight, then $B_1=0$.  
Equivalently, each evaluation point $\alpha_1,\dots,\alpha_r$ is a root of $h$ with multiplicity at least $b$. 
Hence, $h(z) / \prod_{i=1}^r (z- \alpha_i)^b$ is a polynomial of degree at most $t-1-rb$.
Then 
$$
\omega_{C(s-b,r)}\left({\rm ev}_{\alpha_1,\dots, \alpha_r}\left(\frac{h}{ \prod_{i=1}^r (z- \alpha_i)^b}\right)\right) \geq rs-t+1.
$$
Now, by adapting the steps we took for calculating the $U(s,r,b)$-weight of ${\rm ev}_{\alpha_1,\dots, \alpha_r}^{(b)}(f)$ from the minimum weight codeword
of $\Mat_{(s-b)\times r}(\F_q)$ to the calculation of the $U(s,r,b)$-weight of ${\rm ev}_{\alpha_1,\dots, \alpha_r}^{(b)}(h)$ from $h(z) / \prod_{i=1}^r (z- \alpha_i)^b$,
we see at once that 
$$
\omega_{U(s,r,b)}\left({\rm ev}_{\alpha_1,\dots, \alpha_r}^{(b)} (h)\right)  \geq \omega_{U(s,r,b)}\left({\rm ev}_{\alpha_1,\dots, \alpha_r}^{(b)} (f)\right).
$$
Therefore, ${\rm ev}_{\alpha_1,\dots, \alpha_r}^{(b)} (f)$ is indeed a codeword of minimum $U(s,r,b)$-weight in $\Mat_{s\times r}^{(b)}(\F_q)$. 
Since its weight is equal to length-dimension+1, we see that our newly defined code is indeed an MDS code with respect to $U(s,r,b)$-weight. 
This finishes the proof of our theorem.
\end{proof}

\begin{Example}
We consider the bottleneck poset $U(2,3,1)$, whose Hasse diagram is depicted in Figure~\ref{F:U(2,3,1)}.
\begin{figure}[htp]
\begin{center}
\begin{tikzpicture}[scale=.8, every node/.style={circle, draw, fill=white, inner sep=2pt}]
    \node (1) at (0,0) {1};
    \node (2) at (1,1.5) {4};
    \node (3) at (1,0) {2};
    \node (5) at (2,0) {3};
    \draw (1) -- (2);
    \draw (3) -- (2);
    \draw (5) -- (2);
\end{tikzpicture}
\end{center}
\caption{The Hasse diagram of $U(2,3,1)$.}
\label{F:U(2,3,1)}
\end{figure}
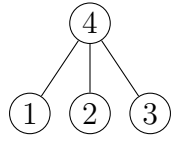 
We explicitly calculate the Reed-Solomon code $ RS^{2,(1)}_{\alpha_1,\alpha_2,\alpha_3}(3) $ in the $ U(2,3,1) $-metric. 
The coefficients are taken from $ \mathbb{F}_5 $, and the evaluation points are $ \alpha_1 = 1 $, $ \alpha_2 = 3 $, and $ \alpha_3 = 4 $. 
The corresponding polynomials, their hyperderivatives, codewords, and weights are tabulated below.

\begin{center}
\adjustbox{max width=\textwidth}{
\begin{tabular}{|c|c|c|c|}
\hline
polynomial & 1-st hyperderivative & codeword & weight \\
\hline 
0 & 0 & $\begin{bmatrix} 0 & 0 & 0 \\ 0 & 0 & 0 \end{bmatrix}$ & 0 \\
\hline
$4x^3 + 3x^2 + x$ & $2x^2 + x + 1$ & $\begin{bmatrix} 3 & 3 & 3 \\ 4 & 2 & 2 \end{bmatrix}$ & 4 \\ 
\hline
$3x^3 + x^2 + 2x$ & $4x^2 + 2x + 2$ & $\begin{bmatrix} 1 & 1 & 1 \\ 3 & 4 & 4 \end{bmatrix}$ & 4 \\
\hline
$2x^3 + 4x^2 + 3x$ & $x^2 + 3x + 3$ & $\begin{bmatrix} 4&4&4 \\ 2&1&1 \end{bmatrix}$ & 4 \\
\hline
$x^3 + 2x^2 + 4x$ & $3x^2 + 4x + 4$ & $\begin{bmatrix} 2 & 2 & 2 \\ 1 & 3 & 3 \end{bmatrix}$ & 4 \\
\hline
1 & 0 & $\begin{bmatrix} 1&1&1 \\ 0 & 0 & 0 \end{bmatrix}$ & 4 \\
\hline
$4x^3 + 3x^2 + x + 1$ & $ 2x^2 + x + 1$ & $\begin{bmatrix} 4&4&4 \\ 4&2&2 \end{bmatrix}$ & 4 \\ 
\hline
$3x^3 + x^2 + 2x + 1$ & $ 4x^2 + 2x + 2$ & $\begin{bmatrix} 2 & 2 & 2 \\ 3  & 4 & 4 \end{bmatrix}$ & 4 \\
\hline
$2x^3 + 4x^2 + 3x + 1$ & $x^2 + 3x + 3$ & $\begin{bmatrix} 0 & 0 & 0 \\ 2  & 1 & 1 \end{bmatrix}$ & 3 \\
\hline
$x^3 + 2x^2 + 4x + 1$ & $3x^2 + 4x + 4$ & $\begin{bmatrix} 3 & 3 & 3\\1  & 3 & 3 \end{bmatrix}$ & 4 \\
\hline
2 & 0 & $\begin{bmatrix} 2&2&2 \\ 0 & 0 & 0 \end{bmatrix}$ & 4 \\
\hline
$4x^3 + 3x^2 + x + 2$ & $ 2x^2 + x + 1$ & $\begin{bmatrix}  0 & 0 & 0 \\ 4 & 2 & 2 \end{bmatrix}$ & 3 \\
\hline

\end{tabular}

\begin{tabular}{|c|c|c|c|}
\hline
polynomial & 1-st hyperderivative & codeword & weight \\
\hline 
$3x^3 + x^2 + 2x + 2$ & $4x^2 + 2x + 2$ & $ \begin{bmatrix} 3 & 3 & 3 \\ 3 & 4 & 4 \end{bmatrix}$ & 4\\
\hline
$ 2x^3 + 4x^2 + 3x + 2$ & $x^2 + 3x + 3$ & $ \begin{bmatrix} 1 & 1 & 1 \\ 2 & 1& 1 \end{bmatrix}$ & 4\\
\hline
$x^3 + 2x^2 + 4x + 2$ & $3x^2 + 4x + 4$ & $ \begin{bmatrix} 4 & 4& 4 \\ 1& 3& 3 \end{bmatrix}$ & 4\\
\hline
3 & 0 & $\begin{bmatrix} 3&3&3 \\ 0 & 0 & 0 \end{bmatrix}$ & 4 \\
\hline
$4x^3 + 3x^2 + x + 3$ & $2x^2 + x + 1$ & $\begin{bmatrix} 1&1&1 \\ 4 & 2 & 2 \end{bmatrix}$ & 4 \\
\hline
$3x^3 + x^2 + 2x + 3$ & $4x^2 + 2x + 2$ & $\begin{bmatrix} 4&4&4 \\ 3 & 4 & 4 \end{bmatrix}$ & 4 \\
\hline
$2x^3 + 4x^2 + 3x + 3$ & $x^2 + 3x + 3$ & $\begin{bmatrix} 2&2&2 \\ 2 & 1 & 1 \end{bmatrix}$ & 4 \\
\hline
$x^3 + 2x^2 + 4x + 3$ & $3x^2 + 4x + 4$ & $\begin{bmatrix} 0&0&0 \\ 1&3&3 \end{bmatrix}$ & 3 \\
\hline
4 & 0 & $\begin{bmatrix} 4&4&4 \\ 0&0&0 \end{bmatrix}$ & 4 \\
\hline
$4x^3 + 3x^2 + x + 4$ & $2x^2 + x + 1$ & $\begin{bmatrix} 2&2&2 \\ 4&2&2 \end{bmatrix}$ & 4 \\
\hline
$3x^3 + x^2 + 2x + 4$ & $4x^2 + 2x + 2$ & $\begin{bmatrix} 0&0&0 \\ 3&4&4 \end{bmatrix}$ & 3 \\
\hline
$ 2x^3 + 4x^2 + 3x + 4$ & $x^2 + 3x + 3$ & $\begin{bmatrix} 3&3&3 \\ 2&1&1 \end{bmatrix}$ & 4 \\
\hline
$x^3 + 2x^2 + 4x + 4$ & $3x^2 + 4x + 4$ & $\begin{bmatrix} 1&1&1 \\ 1&3&3 \end{bmatrix}$ & 4 \\
\hline
\end{tabular}
}
\end{center}

It is easily seen from this table that 
\begin{enumerate}
\item[(1)] The parameters of our code are given by $\text{length} = 4, \text{dimension} = 2$, and $\text{minimum distance} = 3$.
\item[(2)] The $U(2,3,1)$-weight generating polynomial of our code is given by 
$1+4x^3 + 20x^4$.
\end{enumerate}

\end{Example}

\subsection*{Quality Comparison}

In this section, we compare the properties of NRT Reed-Solomon codes and bottleneck Reed-Solomon codes. 
For simplicity, we focus on the bottleneck poset $U(s, r, 0)$, noting that the conclusions of our comparison hold equally for any $b \in \{0, 1, \dots, s-1\}$.

\medskip

The {\em transmission rate} of a code is the ratio $\text{dimension}/\text{length}$. 
The {\em relative distance} of a code is the ratio $\delta := \text{minimum distance}/\text{length}$. 
These two ratios can be used for measuring the data processing and the error correcting capacities of a code. 
We will compare these ratios for the NRT Reed-Solomon codes and the bottleneck Reed-Solomon codes.

We set $b:=0$ and fix three integers $r,s$, and $t$ such that $r\geq 2$ and $1\leq t\leq rs$.
Let $\alpha_1,\dots, \alpha_r$ be $r$ distinct points from $\F_q$.
To ease our notation, we set 
$$
C_1:= RS_{\alpha_1,\dots, \alpha_r}^{s}(t-1)\quad \text{and}\quad C_2:=RS_{\alpha_1,\dots, \alpha_r}^{s,(0)}(t-1).
$$
Hence, $C_1$ is an NRT Reed-Solomon code in ${\Mat}_{s\times r}(\F_q)$ and $C_2$ is a bottleneck Reed-Solomon code in $\Mat_{s\times r}^{(0)} (\F_q)$. 
In this notation, we have the following table of comparisons:

\begin{center}
\resizebox{\textwidth}{!}{%
\begin{tabular}{|c|c|c|c|}
\hline
\textbf{Parameter} & \textbf{$ C_1 $} & \textbf{$ C_2 $} & \textbf{Remarks} \\
\hline
\textbf{Length} & $ n_1 = rs $ & $ n_2 = r(s - 1) + 1 $ & $ C_2 $ is shorter \\
\hline
\textbf{Dimension} & $ k_1 = t $ & $ k_2 = t - r + 1 $ & $ C_1 $ has a higher dimension \\
\hline
\textbf{Minimum Distance} & $ d_1 = rs - t + 1 $ & $ d_2 = rs - t + 1 $ & Both are the same \\
\hline
\textbf{Rate} & $ R_1 = \frac{t}{rs} $ & $ R_2 = \frac{t - r + 1}{r(s - 1) + 1} $ & Depends on $ t $, $ r $, $ s $ values \\
\hline
\textbf{Relative Distance} & $ \delta_1 = \frac{rs - t + 1}{rs} $ & $ \delta_2 = \frac{rs - t + 1}{r(s - 1) + 1} $ & $ C_2 $ has advantage in error correction \\
\hline
\end{tabular}%
}
\end{center}

We conclude from this table that if higher rate and larger dimension are priorities, $C_1$ may be more efficient.
Nevertheless, if robustness to errors per unit length and shorter code length are more important, 
then $C_2$ is more advantageous.
In other words, if maximizing data throughput is critical, $C_1$ could be preferable, while for applications needing robust error correction in a constrained length, $C_2$ might be better suited.
Ultimately, the choice depends on the application context.

\section{AG Codes with Bottleneck Metrics}\label{S:AGinBottleneck}

Let $F/\F_q$ be a global function field with full constant field $\F_q$.
The multiplicative group of nonzero elements of $F$ is denoted by $F^*$. 
For a rational place $P$ of $F$, we denote by $\nu_P$ the corresponding normalized discrete valuation of $F$. 
Let $P_1,\dots, P_r$ be $r$ distinct places. 
For each $i\in \{1,\dots, r\}$, we denote by $t_i$ the local parameter of $P_i$ in $F$ (hence, $\nu_{P_i}(t_i) =1$).
Let $G$ be a divisor of $F/\F_q$. 
For each $i\in \{1,\dots, r\}$, we will denote the coefficient of $P_i$ in $G$ by $n_i$.
The Riemann-Roch space associated with $G$ is given by 
$$
\mc{L}(G) := \{ f\in F^* \mid {\rm div} (f) + G \geq 0 \} \cup \{0\}.
$$
In particular, for $f\in \mc{L}(G)$ and $i\in \{1,\dots, r\}$, if we denote $\nu_{P_i}(f)$ by $-a_i$, then the following inequality holds:
\begin{align}\label{A:definea_i}
-a_i \geq -n_i.
\end{align}
We will make use of the unique local expression for $f$ at $P_i$ ($i\in \{1,\dots, r\}$):
$$
f = \sum_{j=-\infty}^\infty c_{i,j}t_i^j.
$$
Notice that the summation index $j$ in this local expression is shown to vary between $-\infty$ and $\infty$.
However, there is a smallest $j$ such that $c_{i,j}\neq 0$.
It is given by the integer $-a_i$, that is, $\nu_{P_i}(f)$.

For $s\in \Z_+$ and $i\in \{1,\dots, r\}$, let $\mathbf{c}_f^i(s)$ denote the truncated coefficient sequence
\begin{align}\label{A:sthlocalexpressionforf}
\mathbf{c}_f^i(s) := (c_{i, -n_i}, c_{i,-n_i+1},\dots, c_{i, -n_i+s-1}).
\end{align}
We associate to $f$ the following $s\times r$ matrix:
\begin{align*}
f \longmapsto
\mathbf{c}(f) := 
\begin{bmatrix}
c_{1,-n_1} & c_{2,-n_2} & \cdots & c_{r,-n_r} \\ 
c_{1,-n_1+1} & c_{2,-n_2+1} & \cdots & c_{r,-n_r+1} \\ 
\vdots & \vdots & \ddots & \vdots \\
c_{1,-n_1+s-1} & c_{2,-n_2+s-1} & \cdots & c_{r,-n_r+s-1}
\end{bmatrix}.
\end{align*}
It is worthwhile to remark here that the NRT-weight of the $i$-th column of $\mathbf{c}(f)$ is given by 
\begin{align*}
\omega_{\rm C(s,1)} ((\mathbf{c}_f^i(s))^\top) =  s - \min \{ s, a_i + n_i\}.
\end{align*}
The following result is recorded (in a slightly different notation) as Proposition 2 in the reference~\cite{XingNiederreiter2004}.
It serves as a generalization of~\cite[Theorem 6]{RT1997}.

\begin{Proposition}\label{P:byNiederreiterXing}
Let $s\geq 1$ be a positive integer.
Let $G$ be a divisor of $F$ such that $\nu_{P_i}(G) = n_i$ for $i\in \{1,\dots, r\}$. 
If $\deg G$ is in the range $\{g,g+1,\dots, rs-1\}$, then the $\F_q$ vector space $\mathcal{N}$ defined by 
$$
\mathcal{N} := \{ \mathbf{c}(f) \mid f\in \mathcal{L}(G) \} 
$$
satisfies
\begin{itemize}
\item $\dim \mathcal{N} \geq \deg (G) -g+1$,
\item $d_{C(s,r)}(\mathcal{N}) \geq rs-\deg (G)$.
\end{itemize}
\end{Proposition}

\bigskip

We adapt the core idea of the proof of this proposition to the bottleneck metric setting. 
As previously noted, to minimize notational complexity, we will focus exclusively on the case of $U(s,r,0)$. 
We now consider the following subspace of the Riemann-Roch space:
$$
\mathcal{L}^{(0)}_{P_1,\dots, P_r}(G) := \{ f\in \mathcal{L}(G)  \mid  c_{1,-n_1}(f) = \cdots = c_{r,-n_r}(f) \}, 
$$
where $c_{i,-n_i}(f)$ is the coefficient of the smallest degree term in the local expression 
$\sum_{j=-n_i}^\infty c_{i,j} t_i^j$ for $f$ at $P_i$ ($i\in \{1,\dots, r\}$).
The main result of this section is the following. 

\begin{Theorem}\label{T:URTonfunctionfield}
Let $F/\F_q$ be a global function field with genus $g\geq 0$.
Let $r$ and $s$ be two positive integers such that $g \leq r$ and $2\leq s$. 
Let $G$ be a divisor of $F$ such that $\nu_{P_i}(G) = n_i$ for $i\in \{1,\dots, r\}$. 
If the inequalities $$r-1 \leq \deg (G) \leq rs-r+1$$ hold, then the vector space $\mathcal{U}:=\mathcal{U}(G)$ defined by  
$$
\mathcal{U} := \{ \mathbf{c}(f) \mid f\in \mathcal{L}^{(0)}_{P_1,\dots, P_r}(G) \} 
$$
satisfies the following constraints: 
\begin{enumerate}
\item[(1)] $d_{U(s,r,0)}(\mathcal{U}) \geq rs-\deg (G)$,
\item[(2)] $\dim \mathcal{U} \geq \deg (G) -g-r+2$.
\end{enumerate}
\end{Theorem}

\begin{proof}
First of all, the fact that $\mathcal{U}$ is an $\F_q$-vector space does not require a proof. 
We proceed to prove the inequalities (1) and (2).

Let $f\in  \mathcal{L}^{(0)}_{P_1,\dots, P_r}(G) \setminus \{0\}$.
Then the lower bound for the order of the pole of $f$ at $P_i$ (for each $i\in \{1,\dots, r\}$) is $n_i$. 
Let $\alpha$ denote this order defined by  
$$
\alpha = \nu_{P_1}(f)=\cdots = \nu_{P_r}(f).
$$
For $i\in \{1,\dots, r\}$, we set 
$$
w_i(f):= \min \{ s,\alpha+n_i \}.
$$
Since $s\geq 1$ and $f$ is from $\mathcal{L}^{(0)}_{P_1,\dots, P_r}(G) \setminus \{0\}$, $w_i(f)\geq 0$ for every $i\in \{1,\dots, r\}$.
If for some $i\in \{1,\dots, r\}$, $w_i(f)=0$, then $\nu_{P_i}(f) = -n_i$.
In this case, since $f\in  \mathcal{L}^{(0)}_{P_1,\dots, P_r}(G) \setminus \{0\}$, we see that 
$n_1 = \cdots = n_r$. 
Hence, the entire first row of $\mathbf{c}(f)$ is given by the nonzero constant $c_{1,-n_1}$,
implying that 
$$
\omega_{U(s,r,0)}(\mathbf{c}(f)) = rs - r+1= rs - (r-1).
$$
Since our assumption on the degree $G$ is that $\deg (G)\in \{ r-1,r,\dots, rs-r\}$, we see that 
$$
\omega_{U(s,r,0)}(\mathbf{c}(f)) = rs - r+1= rs - (r-1) \geq rs - \deg (G).
$$
In this case, (1) holds.
We proceed with the assumption that $\omega_{U(s,r,0)}(\mathbf{c}(f)) < rs-r+1$. 
In particular, this assumption implies that $0 <  w_i(f)$ and $-n_i < \nu_{P_i}(f)$ for each $i\in \{1,\dots, r\}$.
Therefore, we have 
$$
0 <  w_i(f)-n_i \leq \nu_{P_i}(f)
$$
for every $i\in \{1,\dots, r\}$.
This means that $f$ is contained in the following Riemann-Roch space:
$$
\mathcal{L} \left(G-\sum_{i=1}^r w_i(f)P_i\right).
$$
It follows that the degree of the divisor $G-\sum_{i=1}^r w_i(f)P_i$ is a nonnegative integer,
\begin{align}\label{A:willbestrict}
0 \leq  \deg (G) - \sum_{i=1}^r w_i(f).
\end{align}

We notice also that, since the $U(s,r,0)$-weight of $\mathbf{c}(f)$ is less than $rs-r+1$, 
the NRT weight of $\mathbf{c}(f)$ is equal to the $U(s,r,0)$-weight of $\mathbf{c}(f)$.
But the NRT weight of the $i$-th column of $\mathbf{c}(f)$ is given by $s+ \nu_{P_i}(f)$. 
At the same time, since the first row of $\mathbf{c}(f)$ has no nonzero entries, we see that 
\begin{align*}
\sum_{i=1}^r w_i(f) &= rs - \omega_{\rm NRT} (\mathbf{c}(f)) \\
&= rs - \omega_{U(s,r,0)} (\mathbf{c}(f)) \qquad \text{by Lemma~\ref{L:theyagree}}.
\end{align*} 
We use this information in the degree calculation: 
\begin{align*}
0 &\leq \deg \left(G-\sum_{i=1}^r w_i(f)P_i\right) \\
&=  \deg (G) - \sum_{i=1}^r w_i(f)\\
&= \deg (G) - rs + \omega_{U(s,r,0)} (\mathbf{c}(f)).
\end{align*}
Therefore, we have 
\begin{align*}
rs- \deg (G) \leq  \omega_{U(s,r,0)} (\mathbf{c}(f)).
\end{align*}
This finishes the proof of (1).

We proceed to prove (2). 
Notice that our map $f \mapsto \mathbf{c}(f)$ is essentially the restriction to $\mathcal{L}(G)^{(0)}_{P_1,\dots, P_r}$ 
of the map that defines the NRT-metric code considered in Proposition~\ref{P:byNiederreiterXing}. 
Hence, our map is injective, and the dimension of the image of this map is bounded from below by the dimension of 
$\mathcal{L}(G)^{(0)}_{P_1,\dots, P_r}$. 
But Riemann-Roch theorem implies that $\dim \mathcal{L}(G) \geq \deg(G)-g+1$.
It is not difficult to check that the subspace $\mathcal{L}(G)^{(0)}_{P_1,\dots, P_r}\subset \mathcal{L}(G)$ is defined by the intersection of $r-1$ hyperplanes in $\mathcal{L}(G)$. 
Hence, we see that 
$$
\dim \mathcal{L}(G)\ \geq \ \dim \mathcal{L}(G)^{(0)}_{P_1,\dots, P_r}\ = \ \dim \mathcal{L}(G) -r +1 \geq \deg(G)-g+1 -(r -1).
$$
This finishes the proof of (2).
Hence, the proof of our theorem follows. 
\end{proof}

\begin{Corollary}
We maintain the notation from the previous theorem.
If the genus of $F$ is 0, then $\mathcal{U}$ is an MDS code in the bottleneck metric. 
If the genus of $F$ is 1, then $\mathcal{U}$ is either an MDS bottleneck metric code or a near MDS bottleneck metric code.
\end{Corollary}
\begin{proof}
By adding the inequalities (1) and (2) of our previous theorem, we obtain 
$$
\dim \mathcal{U}+ d_{U(s,r,0)}(\mathcal{U}) \geq rs-\deg (G) + \deg (G) -g-r+2 =rs- r +1 + (1-g).
$$
If $g=0$, then $rs-r+2$ is bigger than the length of $\mathcal{U}$.
Hence, Singleton's upper bound is achieved. 

We proceed with the assumption that $g=1$. 
Then the dimension plus the minimum distance of $\mathcal{U}$ is at least $rs-r+1$.
It follows that $\mathcal{U}$ is either an MDS code or a near-MDS code in the bottleneck metric. 
\end{proof}

In the next section, we will improve our previous theorem. 

\subsection*{MDS AG Bottleneck Metric Codes}

Our goal in this section is to show that if we choose our function field carefully, we can always construct MDS bottleneck metric codes. 
In this section, we prove the second main result of our paper, that is, Theorem~\ref{T:MDSURT}.
We recall its statement for convenience.
\medskip

Let $F/\F_q$ be a global function field of genus $g\geq 1$.
Let $r$, $k$ and $s$ be positive integers satisfying the inequalities 
\begin{itemize}
\item $2r+s-g\leq rs$,
\item $0\leq k-1\leq g\leq r$, and 
\item $g-1\leq s$.
\end{itemize}
We assume that $F/\F_q$ has at least $r$ rational places.
Let $h$ denote the divisor class number of $F$. 
Finally, let $A_k$ denote the number of positive divisors of $F$ degree $k$. 
If the inequality 
\begin{align}\label{A:keyinequality}
{ r+s + k - g \choose r-1 } A_k < h
\end{align}
holds, then the parameters of the AG code defined by $\mathcal{U} := \{ \mathbf{c}(f) \mid f\in \mathcal{L}^{(0)}_{P_1,\dots, P_r}(G) \}$
satisfy the following inequalities 
\begin{enumerate}
\item[(1)] $d_{U(s,r,0)}(\mathcal{U}) \geq s-g+2+k$,  
\item[(2)] $\dim \mathcal{U} \geq rs - r +1 -s$.
\end{enumerate}

\begin{proof}[Proof of Theorem~\ref{T:MDSURT}.]
Our proof is very similar to the proof of~\cite[Theorem 3]{XingNiederreiter2004}.
Let $P_1,\dots, P_r$ be $r$ mutually distinct rational places of $F$.

We consider the divisors of $F$ of the form 
\begin{align}\label{A:Dplus}
D+ \sum_{i=1}^r u_i P_i
\end{align}
where $D$ is a positive divisor of degree $k$ and $0\leq u_i \leq s-2$ for $i\in \{1,\dots, r\}$ such that 
\begin{align}\label{A:u_i's}
\sum_{i=1}^r u_i = (rs - r) - s - (-g+k +1) = rs-r-s+g-k-1.
\end{align}
Note that $- s - (-g+k +1)\leq 0$. 
The number of such divisors is bounded from above by ${ r-1+s - g+k+1 \choose r-1 } A_k $.
By our assumption, this number is less than $h$, the divisor class number of $F$. 
But for every degree, there are exactly $h$ divisor class of that degree. 
Hence, our assumption implies that there exists a divisor $G$ such that $\deg (G) = rs - r -s+ g-1$
and $G$ is not linearly equivalent to any of the divisors defined in (\ref{A:Dplus}).
For this $G$, we claim that the Riemann-Roch space
$$
\mathcal{L}\left( G - \sum_{i=1}^s u_i P_i \right)
$$
is trivial as long as the $u_i$'s satisfy (\ref{A:u_i's}).
Towards a contradiction, assume that there exists a nonzero rational function $g\in \mathcal{L}\left( G - \sum_{i=1}^s u_i P_i \right)$. 
Then the divisor $E$ defined by 
$$
E:= {\rm div}(g)+ G - \sum_{i=1}^s u_i P_i
$$
is positive, and its degree is $k$. 
Since the definition of $E$ implies that $$G = E+\sum_{i=1}^s u_i P_i - {\rm div}(g),$$ we see that 
$G$ is linearly equivalent to $E+\sum_{i=1}^s u_i P_i$. 
But this contradicts with our choice of $G$ above. 

Now, our assumptions on $r,s$, and $g$ give us the inequalities 
$$
r-1\ \leq \ rs-r-s+g-1\ \leq \ rs-r.
$$
Hence, our divisor $G$ satisfies the hypothesis of Theorem~\ref{T:URTonfunctionfield}. 
Then the associated vector space $\mathcal{U}:=\mathcal{U}(G)$ has parameters satisfying the inequalities in Theorem~\ref{T:URTonfunctionfield}.
Nonetheless, there is an improvement in the minimum distance. 
Indeed, let $f$ be a nonzero function in $\mathcal{L}(G)$.
Then we know from the proof of Theorem~\ref{T:URTonfunctionfield} that  
$$
\omega_{U(s,r,0)} (\mathbf{c}(f)) = rs-r+1 - \sum_{i=1}^r w_i(f)
$$
and 
$$
f\in \mathcal{L}\left( G - \sum_{i=1}^r w_i(f) P_i\right).
$$
But (\ref{A:u_i's}) implies that 
$$
\sum_{i=1}^r w_i(f) \leq rs-r-s+g-k-1.
$$
Hence, the minimum distance has a bigger lower bound as the following calculation shows: 
$$
\omega_{U(s,r,0)}(\mathcal{U}) \geq (rs-r+1) - (rs-r-s+g-k-1) = s-g+k+2.
$$
This finishes the proof of our theorem.
\end{proof}

\begin{Corollary}\label{C:directconsequence}
We maintain the notation from the previous theorem.
If the genus of $F$ is $k-1$, then $\mathcal{U}$ is an MDS bottleneck metric code. 
In particular, if $F$ is the function field of an elliptic curve, then $\mathcal{U}$ is either an MDS bottleneck-metric code or a near MDS bottleneck metric code.
\end{Corollary}
\begin{proof}
By adding the inequalities (1) and (2) of our previous theorem, we obtain 
$$
\dim \mathcal{U}+ d_{U(s,r,0)}(\mathcal{U}) \geq rs- r +1 + (2+k-g).
$$
If $g=k-1$, then the lower bound becomes $rs-r+2$.
Hence, the Singleton's upper bound is achieved. 
This finishes the proof of our first assertion. 
For our second assertion, we observe that if $g=1$, then $k$ is necessarily 0.
Hence, we see that $\mathcal{U}$ is an MDS code by the first assertion. 
This finishes the proof of our corollary. 
\end{proof}

\section*{Acknowledgements}

The first two authors thank the Louisiana Board of Regents for their support through the grant LEQSF(2023-25)-RD-A-21.

\bibliographystyle{plain}
\bibliography{references}

\end{document}